\documentclass{aa}
\usepackage{longtable}

\usepackage{psfig}
\usepackage{supertabular}

\begin{document}
%
\def\lfir{$L_{\rm FIR}$}
\def\mabs{M$_{\rm abs}$}
\def\etal{et al.}
\def\hii{H{\sc ii}}
\def\cbeta{$c_{\rm H\beta}$}
\def\av{A$_{\rm v}$}
\def\flam{$F_{\lambda}$}
\def\ilam{$I_{\lambda}$}
\def\micron{$\mu$m}
\def\mum{$\mu$m}
\def\kms{km s$^{-1}$}
\def\kmsmpc{km s$^{-1}$ Mpc$^{-1}$}
\def\cmc{cm$^{-3}$}
\def\erg{ergs s$^{-1}$ cm$^{-2}$ \AA$^{-1}$}
\def\ergs{ergs s$^{-1}$}
\def\ergscm{ergs s$^{-1}$ cm$^{-2}$}
\def\lsun{L$_{\odot}$}
\def\msun{M$_{\odot}$}
\def\zsun{Z$_{\odot}$}
%
\def\halpha{\ifmmode {\rm H{\alpha}} \else $\rm H{\alpha}$\fi}
\def\hbeta{\ifmmode {\rm H{\beta}} \else $\rm H{\beta}$\fi}
%
%
\def\heia{He\,{\sc i} $\lambda$4471}
\def\heib{He\,{\sc i} $\lambda$4922}
\def\heic{He\,{\sc i} $\lambda$5876}
\def\heiia{He\,{\sc ii} $\lambda$4686}
\def\heiib{He\,{\sc ii} $\lambda$5412}
\def\heii{\ifmmode {{\rm He{\sc ii}} \lambda 4686} \else {He\,{\sc ii} $\lambda$4686}\fi}
\def\Heii{He~{\sc ii}}
\def\Hei{He~{\sc i}}

%
\def\oia{[O\,{\sc i}] $\lambda$6300}
\def\oib{[O\,{\sc i}] $\lambda$6364}
\def\oii{[O\,{\sc ii}] $\lambda$3727}
\def\oiii{[O\,{\sc iii}]}
\def\oiiia{[O\,{\sc iii}] $\lambda$4959}
\def\oiiib{[O\,{\sc iii}] $\lambda$5007}
\def\ov{O\,{\sc v} $\lambda$5590}
%
%
\def\ni{[N\,{\sc i}] $\lambda$5199}
\def\nii{[N\,{\sc ii}] $\lambda$5755}
\def\niii{N\,{\sc iii} $\lambda$4640}
\def\nv{N\,{\sc v} $\lambda$4604}
%
%
\def\ciii{C\,{\sc iii} $\lambda$4650}
\def\ciiia{C\,{\sc iii} $\lambda$4658}
\def\ciiib{C\,{\sc iii} $\lambda$5696}
\def\civ{C\,{\sc iv} $\lambda$5808}
%
%
\def\sii{[S\,{\sc ii}]}
\def\siia{[S\,{\sc ii}] $\lambda$6716}
\def\siib{[S\,{\sc ii}] $\lambda$6731}
\def\Sii{[S\,{\sc ii}] $\lambda\lambda$6716,6731}
\def\siii{[S\,{\sc iii}] $\lambda$6318}
%
%
\def\ariva{[Ar\,{\sc iv}] $\lambda$4711}
\def\arivb{[Ar\,{\sc iv}] $\lambda$4740}
%
%
\def\feiiia{[Fe\,{\sc iii}] $\lambda$4658}
\def\feiiib{[Fe\,{\sc iii}] $\lambda$5271}
%
%
\def\cliiia{[Cl\,{\sc iii}] $\lambda$5518}
\def\cliiib{[Cl\,{\sc iii}] $\lambda$5538}
%
%
\def\oiiishb{[O\,{\sc iii}]/H$\beta$}
\def\niisha{[N\,{\sc ii}]/H$\alpha$}
\def\siisha{[S\,{\sc ii}]/H$\alpha$}

\def\Siii{[S~{\sc iii}]}
\def\Siv{[S~{\sc iv}]}
\def\Oii{[O~{\sc ii}]}
\def\Oiii{[O~{\sc iii}]}
\def\Oiv{[O~{\sc iv}]}
\def\Neii{[Ne~{\sc ii}]}
\def\Neiii{[Ne~{\sc iii}]}
\def\Nev{[Ne~{\sc v}]}
\def\Arii{[Ar~{\sc ii}]}
\def\Ariii{[Ar~{\sc iii}]}

\def\brg{Br$\gamma$}
\def\bra{Br$\alpha$}
\def\arii{[Ar~{\sc ii}] 6.99 $\mu$m }
\def\pfa{Pf$\alpha$}
\def\ariii{[Ar~{\sc iii}] 8.99 $\mu$m }
\def\siv{[S~{\sc iv}] 10.51 $\mu$m }
\def\neii{[Ne~{\sc ii}] 12.81 $\mu$m}
\def\nev{[Ne~{\sc v}] 14.3 $\mu$m}
\def\neiii{[Ne~{\sc iii}] 15.55 $\mu$m}
\def\siii{[S~{\sc iii}] 18.71 $\mu$m}
\def\oiv{[O~{\sc iv}] 25.9 $\mu$m}
\def\SIii{[Si~{\sc ii}] 34.81 $\mu$m}
\def\oivrat{[O~{\sc iv}]/([Ne~{\sc ii}]+0.44[Ne~{\sc iii}])}

\def\Q0{\ifmmode {Q_{{\rm H}^0}} \else {$Q_{{\rm H}^0}$}\fi}

%
\def\aap{A\&A}
\def\aas{A\&AS}
\def\aj{AJ}
\def\apj{ApJ}
\def\apjl{ApJ}
\def\apjs{ApJS}
\def\mnras{MNRAS}
\def\pasp{PASP}



\thesaurus{11(11.19.3; 11.19.5; 09.08.1; 08.23.2; 13.09.1; 11.09.2)} 

\title{On the origin of \Oiv\ emission in Wolf-Rayet galaxies
}

\author{Daniel Schaerer \inst{1} \and
           Gra\.{z}yna Stasi\'nska\inst{2} }

\offprints{D. Schaerer}

\institute{
Laboratoire d'Astrophysique, Observatoire Midi-Pyr\'en\'ees, 14, Av. E. Belin, 
F-31400 Toulouse, France (schaerer@obs-mip.fr)
\and
DAEC, Observatoire de Meudon, Meudon, France (grazyna@obspm.fr)
}

\date{Received 9 march 1999 / Accepted 29 march 1999}

\authorrunning{Schaerer \& Stasi\'nska}
\maketitle

\begin{abstract} 
We propose that the emission of the high excitation \oiv\ line observed
with ISO in \object{NGC 5253} and \object{II Zw 40} is due to the presence of hot Wolf-Rayet
(WR) stars in these objects. 
We construct a consistent evolutionary synthesis and photoionization model 
which successfully reproduces the constraints on their massive star content 
and the relevant optical and IR emission lines including \oiv.
Our explanation for the origin of \Oiv\ is supported empirically by: 
{\em 1)} the simultaneous presence of nebular \Heii\ and \Oiv\ in these 
objects, and 
{\em 2)} the close relation between nebular \Heii\ and WR stars in 
extragalactic \hii\ regions.
Photoionization by hot WR stars is mainly expected to be of importance
in young low metallicity galaxies.
Alternate mechanisms are likely at the origin of \oiv\ emission 
in other objects.

\keywords{Galaxies: starburst -- Galaxies: stellar content -- \hii\ regions 
        -- Stars: Wolf-Rayet -- Infrared: galaxies -- Galaxies: individual:
        NGC 5253, II Zw 40}

\end{abstract}

%

\section{Introduction}

Infrared observations of emission line galaxies give access to ions 
not visible in the optical domain. 
Among those is the high excitation \oiv\ line which has not only been 
observed in active galactic nuclei but also in several starburst galaxies, 
although at a much fainter level (Genzel \etal\ 1998, Lutz \etal\
1998, hereafter LKST98).

Hot, massive stars generally emit only few ionizing photons with 
energies above the  \Heii\ edge at 54.42 eV required for the production 
of \Oiv.
Therefore the origin of this high excitation line in starbursts 
has been unclear so far. Different excitation mechanisms 
(weak AGNs, super-hot stars, planetary nebulae, and ionizing shocks)
have been discussed by LKST98.
Based on simple estimates, photoionization and shock models, these authors
favor 
ionizing shocks related to the starburst
activity as  the most likely explanation for \oiv\ emission {\em in general}.

According to LKST98, massive super-hot stars remain, however, an option 
for the high excitation dwarf galaxies included in their sample.
One of these objects (NGC 5253) was studied in more detail by Crowther \etal\
(1999, hereafter C99), who showed, by computing photoionization models
around single stars, that WNE-w stars can indeed produce strong \Oiv\ and 
\Nev\ in surrounding \hii\ regions. 
The fact that these lines were not prominent in NGC 5253 led them to 
exclude the possibility of a significant number of such 
stars being present in this galaxy. 
As will be shown below we do not support their conclusion for a variety 
of reasons.

To shed more light on the origin of the \oiv\ emission in dwarf 
galaxies, we use  the information on nebular properties
and stellar content derived from optical studies to complement the 
information from IR data. In addition to 
NGC 5253, we also consider II Zw 40. These are the two  compact low metallicity
 galaxies which show the highest excitation among the 
starbursts observed by LKST98.
Both objects are known as so-called WR galaxies
(cf.\ Conti 1991, Schaerer \etal\ 1999b), where the presence
of broad stellar emission testifying to 
the presence of WR stars
provides powerful constraints on the burst age and massive star
content (e.g.\ Schaerer et al.\ 1999a).

In this paper, we present a stellar population model which reproduces 
the observed stellar features and which, used as an input for 
photoionization models, explains at the same time the ionization 
structure of the nebular gas as revealed by the optical and IR 
fine structure lines. 

\section{On the association of \oiv\ with \Heii}
\label{s_ass}
In the sample of \Oiv\ emitting starbursts of LKST98, II Zw 40 and NGC 5253
show the strongest excitation (measured 
by \Neiii/\Neii), the largest \Oiv\ strength (quantified by \oivrat;
cf.\ LKST98, Fig.\ 2) and stand out by several properties:

\begin{itemize}
\item[{\em 1)}] 
 Nebular \heii\ emission indicative of high excitation is present
in the region dominating the optical emission 
(Walsh \& Roy 1989, 1993, hereafter WR89, WR93, Guseva \etal\ 1998)

\item[{\em 2)}] 
A significant number of Wolf-Rayet stars has been detected in these
regions (Kunth \& Schild 1981, Walsh \& Roy 1987, Vacca \& Conti 1992, 
Schaerer \etal\ 1997, hereafter SCKM97)

\item[{\em 3)}]
In both the dwarf II Zw 40 and the amorphous galaxy NGC 5253 one or few
star forming regions of a young age clearly dominate the production of
ionizing photons (Vanzi \etal\ 1996, Beck \etal\ 1996, Calzetti \etal\ 1997).
\end{itemize}

Finding 1) confirms the presence of high energy photons ($>$ 54 eV) 
deduced from the IR observations of \Oiv\ and naturally suggests a direct 
link between the nebular \Heii\ and \Oiv\ emission. 
Furthermore the optical observations allow a more precise localisation
of the high excitation regions.
1) and 2) indicate that II Zw 40 and NGC 5253
are objects where the observed \Heii\ emission is likely due to hot WR stars
(Schaerer 1996, 1997, 1998; De Mello \etal\ 1998).
3) justifies, at least to first order, the use of a ``spectral 
template'' of the brightest starburst region as a representation of the 
ionizing spectrum of the entire region covered by the ISO observations.

\section{Stellar population}
\begin{figure}[htb]        
\centerline{\psfig{figure=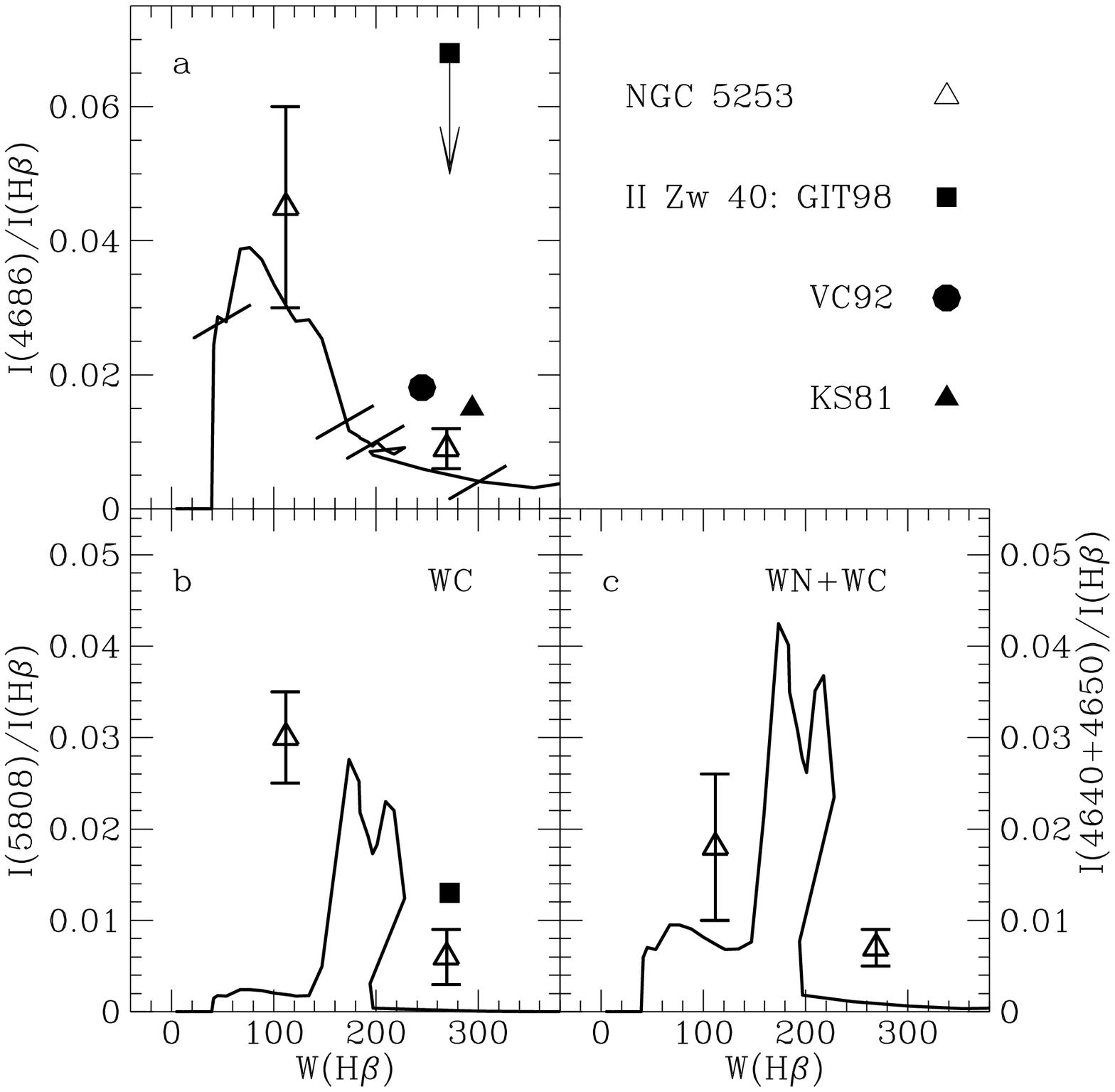,width=8.8cm}
}
\caption{Observed and predicted WR lines intensities as a function
of $W(\hbeta)$:
broad \heii\ ({\bf a}), \civ\ ({\bf b}), and \niii + \ciii\ ({\bf c}).
Model predictions for an instantaneous burst with a Salpeter IMF
for Z=0.004. Observational data from: NGC5253 -- regions A and B from 
Schaerer \etal\ (1999a), II Zw 40 -- data from Kunth \& Schild (1981), 
Vacca \& Conti (1992) and Guseva \etal\ (1998).
Tickmarks in a) represent ages of 2.8, 3.0, 4.2, and 5.0 Myr respectively.
The observed WR and O star populations are well reproduced for
burst ages of $\sim$ 3-5 Myr.
}
\label{fig_ihb}
\end{figure}

WR stars of both WN and WC types have been observed in the two dominant
regions of NGC 5253 by SCKM97. 
In II Zw 40 broad \heii\ indicative of WR stars was detected by
Kunth \& Sargent (1981), Vacca \& Conti (1992) and Guseva \etal\ (1998).
The latter also detect broad \civ\ emission due to WC stars.
The WR and O star content was already analysed by Schaerer (1996), 
SCKM97 and Schaerer \etal\ (1999a).
In Fig.\ \ref{fig_ihb} the observed intensities of the various WR features 
are shown and compared to an instantaneous burst model  of Schaerer \& 
Vacca (1998) with a Salpeter IMF at the appropriate metallicity 
(Z/\zsun $\sim$ 1/5). 
The observations of Guseva \etal\ (1998) refer to the entire
``WR bump'' (4643-4723 \AA) and represent therefore an upper limit.
Shifts in $W(\hbeta)$ of the theoretical predictions for the WC lines with respect
to the observations are not significant since they correspond to
very short timescales. This and other potential uncertainties 
affecting such a comparison have been extensively discussed in Schaerer \etal\
(1999a). 
Figure \ref{fig_ihb} shows that all line strengths are reasonably well 
reproduced by the model. 
At the corresponding ages of $\sim$ 3-5 Myr (cf.\ SCKM97) our synthesis model
provides therefore a good description of the massive star content
in these regions.

\section{Photoionization models}
The spectral energy distributions predicted by the synthesis models 
described above have been used as input to the photoionization 
code {\em PHOTO} (same version as in Stasi\'nska \& Leitherer 1996, 
hereafter SL96). The remaining input parameters are the total 
number of stars, the density distribution and the chemical 
composition here taken as Z/\zsun=1/4 for easy comparison with SL96
(the ionization structure of a nebula is insensitive to a small 
change in the abundances in the gas). 
Following SL96 we calculate sequences of models for a 
spherical gaz distribution with a uniform hydrogen density $n$ and 
filling factor $\epsilon$, both assumed 
constant during the evolution of the starburst.  
For a given age, models with the same ionization parameter
$U = A(\Q0 n \epsilon^2)^{1/3}$ have the same ionization structure.
\Q0\ is the total number of photons above 13.6 eV, and $A$ a 
function of the electron temperature (see SL96).
The densities derived in II Zw 40 and NGC 5253 ($\sim$ 70--300
cm$^{-3}$, WR89, WR93, C99) are low enough, so that collisional 
deexcitation is negligible for the lines of interest.
We therefore explore the parameter space by simply taking $n$=10 cm$^{-3}$ and 
$\epsilon$=1, and consider three different initial masses for the 
starburst: $10^{3}$, $10^{6}$ and $10^{9}$ \msun.
These three model sequences will be referred to as the 
sequence with low, intermediate and high $U$.

\begin{figure}[htb]        
\centerline{
\psfig{figure=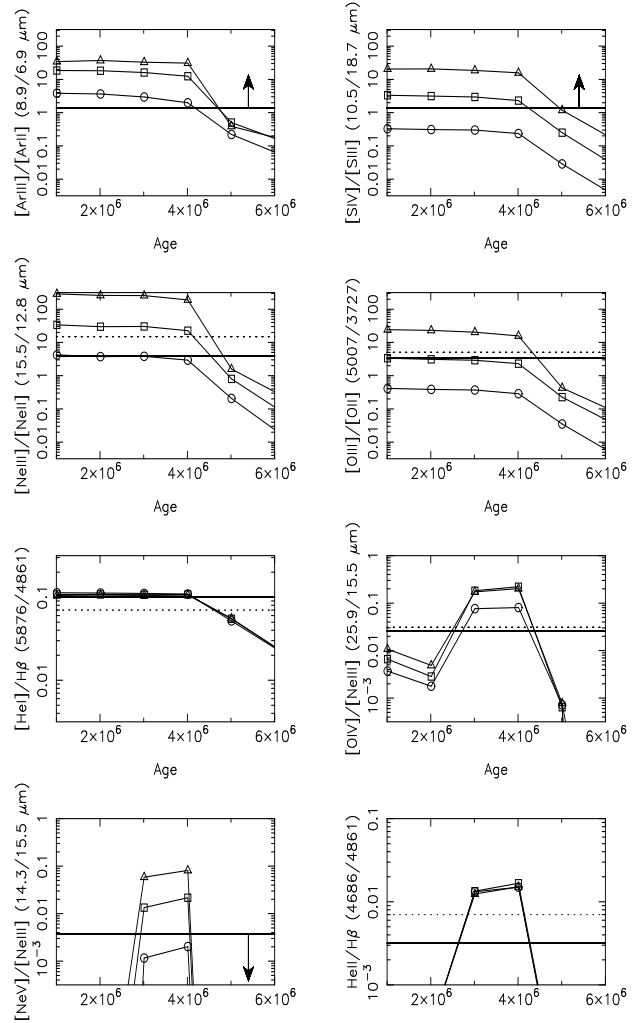,width=8.8cm,height=14cm}
}
\caption{Comparison of observed and predicted IR and optical line 
ratios for instantaneous burst models as a function of time.
{\bf Models:}
Circles denote low $U$ models, squares intermediate $U$, 
triangles high $U$ models. Filled spheres with $n=10$ cm$^{-3}$.
{\bf Observations:}
NGC 5253 -- solid lines, II Zw 40 -- dotted lines.
Sources given in text.
}
\label{fig_ss_1}
\end{figure}

In Fig.\ \ref{fig_ss_1} we show the temporal evolution of selected  
line ratios. 
Unlike C99, we chose to show line ratios that are independent of the 
abundances of the parent elements, in order to facilitate comparison with 
observations of different galaxies. The only exceptions are \Oiv/\Neiii\
\footnote{The O/Ne ratio is 5.0 in the models compared to 4.9 -- 5.5 
in II Zw 40 (WR93) and 4.--7.4  in NGC 5253 (WR89).}
and \heic/\hbeta\ when helium is fully ionized in the \hii\ region.
Also, we limit ourselves to line ratios that involve a dominant ionic 
stage in the nebula.
Line ratios like \oiv/\neii\ or \nev/\neii\ as used by Genzel et al. 
(1998) are difficult to interpret, as the lines are likely emitted by 
very different regions.

As expected, line ratios from adjacent ionic stages show a progressive 
decrease of the overall excitation with time for the models we are considering.
Helium remains fully ionized up to 4 Myr.
Notable exceptions to this trend are \Oiv, \Nev\ and \Heii, species 
with the ionization potential at or above the \Heii\ edge, which appear 
during a short phase (at ages $t \sim$ 3-4 Myr) where hot WR stars 
provide a non-negligible flux above 54 eV (see Schaerer \& Vacca 1998).
Slower temporal changes would of course be obtained for
non instantaneous bursts. 

It must be noted that those line ratios which are function of the ionization 
parameter also depend somewhat on the adopted geometry, and 
this in a non trivial way. 
For example, in models with a thin shell geometry, the line ratios shown here
differ by factors up to 3 from models for full spheres with the same mean 
ionization parameter. 
%
In the case of the \Nev/\Neiii\ ratio, the value predicted during the WR 
phase can be smaller by a factor of about 10 since, despite the presence
of high energy photons,
there is no matter emitting at a high ionization parameter, close to the 
star cluster. 

In the following we compare the model predictions
with observations of NGC 5253 and II Zw 40.

\section{Comparison with observations of NGC 5253 and II Zw 40}
\label{s_obs}
Observed line ratios are overplotted on the model predictions
shown in Fig.\ \ref{fig_ss_1}.
The optical data is taken from WR89 and WR93 (region 1 in both 
objects).
IR fluxes for NGC 5253 are taken from Genzel \etal\ (1998), LKST98,
and C99. 
For  \Siv/\Siii\ a lower limit is obtained since different
ISO apertures are involved in the measurements.
All other limits are  ``real'' detection limits.
Adopting the Draine (1989) extinction curve and $A_v=7.7$ mag
(C99) increases the \Ariii/\Arii\ and \Siv/\Siii\ ratios by $\sim$
40 \%. Other IR line ratios are much less affected.
C99 consider two separate emission regions to be responsible 
for the high excitation lines (e.g.\ \Siv) and lower excitation lines
respectively (e.g.\ \Neii). We see no compelling reason for such 
a somewhat ``artificial'' separation. A similar structure is e.g.\ 
naturally obtained in an ideal spherical nebula.
Instead of using line fluxes corrected for such effects we therefore
use the original measurements.
IR line ratios for II Zw 40 are from LKST98. We have no access to 
the acquired ISO SWS spectra, which should, however, become 
available soon. 

From the top panels of Figure \ref{fig_ss_1} it is evident that no 
single 
model can reproduce at the same time all the 
observed line ratios.
This finding is not surprising and may be due to several reasons:
1) The structure of the galaxies is more complicated than assumed 
in the models.
2) Although one or few bursts of similarly young age dominate the 
ionizing flux (Sect.\ \ref{s_ass}), the ionizing spectrum is likely 
not fully described by a single burst population.
3) Atomic data may be inaccurate. In particular, the computation of
collision strenghts for fine structure transitions 
is a very delicate problem, and the evaluation of the 
formal uncertainty is difficult. A comparison between 
the plasma diagnostics obtained using different IR and optical 
lines for the planetary nebula \object{NGC 6302} led Oliva 
\etal\ (1996) to suggest that the collision strengths which enter in 
the calculation of the intensity of \nev\ are overestimated by a factor 3! 
Similar problems are likely to occur for other fine structure 
lines.

The \Oiii/\Oii\ ratio, which is one of the best studied from all 
points of view, indicates that, if the age of the starburst lies 
between 3 and 4 Myr, as indicated by the Wolf-Rayet features, the 
models with intermediate $U$ are the most adequate to represent the 
two galaxies under study. At such an age, helium is still completely 
ionized, because the radiation field is hard enough. 
The discrepancy with the measurement of \heic/\hbeta\ in II Zw 40
is likely due to absorption by Galactic interstellar sodium intervening
at this redshift (Izotov 1999, private communication).

Our main result is illustrated in the last three panels of Fig.\ 
\ref{fig_ss_1} where we show that during a short phase the stellar 
population provides enough photons above 54.4 eV to naturally produce 
the \heii, \Oiv\ and \Nev\ lines at levels comparable to the 
observed ones.
The emission is due to the presence of hot WR stars at ages
$t \sim$ 3-4 Myr (cf.\ Schaerer 1996, Schaerer \& Vacca 1998).
The predicted strength of these lines exceeds even somewhat 
the observations\footnote{The observations of Guseva \etal\ (1998) 
indicate a larger value \heii/\hbeta =0.018 for II Zw 40.}.
However, this does not invalidate our conclusion.
A more realistic population ``mix'' can easily 
reconcile the intensity of  \heii\ with the observations. 
As for the \Oiv\ and \Nev\ lines, 
they are sensitive to the geometry (see above), which provides 
ample space for fitting with tailored photoionization models.
This should, however, only be undertaken when the relevant atomic data
have been validated by detailed multiwavelength studies of simpler objects
and by comparisons with photoionization models.

\section{Summary and discussion}
We propose that \Oiv\ 25.9 \mum\ emission in NGC 5253 and II Zw 40 is 
due to the presence of hot WR stars observed in both objects.
We draw this conclusion from both empirical and theoretical facts.
First, we note that nebular \heii\ and \Oiv\ emission occur simultaneously in these
objects. Furthermore a close link between nebular \Heii\ and WR stars
has now been established for the so-called WR galaxies, 
extra-galactic \hii\ regions and the few Local Group \hii\ regions 
exhibiting this feature
(Schaerer 1996, 1997, 1998; see Schaerer \etal\ 1999b for a catalogue
of these objects).
Second, quantitative models of the stellar populations using up-to-date
non-LTE atmospheres including stellar winds and the most recent evolutionary 
tracks are able to explain the observed massive star population
and the optical emission lines (including nebular \heii) during the 
WR rich phase (Schaerer 1996, 1998), even for the lowest metallicity
object, \object{I Zw 18} (De Mello \etal\ 1998, Stasi\'nska \& Schaerer 1999),
where such stars were detected recently.
In addition, as demonstrated here, the \Oiv\ emission (and other IR
lines) is also naturally reproduced by these models.
The two objects considered here are the best available to constrain
the origin of \Oiv\ emission: they show the highest excitation,
and represent the most simple objects in terms of their ionizing
population.

IR observations of the 8 Local Group (LG) \hii\ regions known to 
exhibit nebular \Heii\ (cf.\ Garnett \etal\ 1991) can provide
a simple ``consistency'' test: the presence of this line is 
a necessary condition for showing \Oiv\ emission. The presence
or absence of \oiv\ is, however, also influenced by the ionization
parameter and the nebular geometry (cf.\ above).

Observational evidence suggests that such high excitation \hii\ regions
occur preferentially at low metallicities. This holds both for the LG 
and extragalactic objects (cf.\ Schaerer 1997, 1998), including 
II Zw 40 and NGC 5253. The same can thus be expected for the contribution
of WR stars to \Oiv.
Low metallicity may indeed also justify the neglect of line blanketing 
in the WR models of Schmutz \etal\ (1992) included in our synthesis 
models. The effects discussed by C99 and Crowther (1998) suppressing 
the output of photons above the \Heii\ edge in metal-rich WR models 
and/or high density winds could well be ineffective at low metallicities,
as also suggested by the empirical evidence.

Our explanation for the stellar photoionization origin of \Oiv\
in dwarf-like low metallicity galaxies cannot be necessarily 
generalised to all the objects of LKST98. 
Although indeed 6 out of 14 from their list are known WR ``galaxies''
(Schaerer \etal\ 1999b), it is unlikely that the regions where WR 
stars are detected contribute a significant fraction of the total 
ionizing flux in these complex objects.
Outflows, weak Seyfert activity, and other phenomena are known in
some of them and provide alternative explanations as discussed
by LKST98.
More complex models will be required to interpret such objects, 
to provide an theoretical understanding of new empirical IR
diagnostic diagrams (cf.\ Genzel \etal\ 1998), and to assess the
contribution of stellar sources to high energy photons.

\begin{acknowledgements}
Yuri Izotov kindly provided us with data prior to publication.
We thank Marc Sauvage and Suzanne Madden for useful discussions.
DS acknowledges a grant from the Swiss National Foundation of Scientific
Research.

\end{acknowledgements}



\begin{thebibliography}{}


\bibitem[\protect\astroncite{Calzetti \etal}{1997}]{CALZETTIetal97}
Calzetti D., Meurer G.R., Bohlin R.C., Garnett D.R., Kinney A.L., 
Leitherer C., \& Storchi-Bergmann T., 1997, AJ 114, 1834

\bibitem[]{}
Conti, P.S., 1991, \apj, 377, 115

\bibitem[]{} Crowther, P.A.C., 1998
in ``Wolf-Rayet Phenomena in Massive Stars and Starburst Galaxies'', 
        K.A. van der Hucht, G. Koenigsberger, \& P.R.J. Eenens (eds.), 
        IAU Symp. 193, (San Francisco: ASP), in press (astro-ph/9812402)

\bibitem[]{}
Crowther, P.A.C., Beck, S.C., Willis, A.J., Conti, P.S., Morris, P.W., Sutherland,
R.S., 1999, \mnras, in press (astro-ph/9812080; C99)

\bibitem[]{}  De Mello, D., Schaerer, D., Leitherer, C., Heldmann, J., 1998,
\apj, 507, 199

\bibitem[]{}
Draine, B.T., 1989, in ``Infrared Spectroscopy in Astronomy'', Ed. B.H. Kaldeich,
ESA SP-290, 93 

\bibitem[]{}
Garnett, D.R., Kennicutt, R.C., Chu, Y.H., Skillman, E.D.
1991, \apj, 373, 458

\bibitem[]{}
Genzel, R., \etal, 1998, \apj, 498, 579

\bibitem[]{}
Guseva, N.G., Izotov, Y.I., Thuan, T.X., 1998, in preparation

\bibitem[]{}
Kunth, D., Sargent, W.L.W., 1981, \aap, 101, L5


\bibitem[]{}
Izotov, Y.I., Thuan T.X., 1998, \apj, 227, 237


\bibitem[]{}
Lutz, D., \etal, 1996, \aap, 315, L17

\bibitem[]{}
Lutz, D., Kunze, D., Spoon, H.W.W., Thornley, M.D., 1998, \aap, 333, L75
(LKST98)


\bibitem[Oliva Pasquali & Reconditi 1996]{1996A&A...305L..21O} Oliva, E., 
Pasquali, A., Reconditi, M. 1996, \aap, 305, L21
   

\bibitem[Schaerer 1996]{1996ApJ...467L..17S} Schaerer, D. 1996, \apjl, 467, 
L17

\bibitem[Schaerer 1997]{} Schaerer, D., 1997, in ``Dwarf Galaxies: Probes for
Galaxy Formation and Evolution'', Ed. J.Andersen, Highlights in Astronomy, in press
(astro-ph/9812345)

\bibitem[]{} Schaerer, D., 1998,
in ``Wolf-Rayet Phenomena in Massive Stars and Starburst Galaxies'', 
        K.A. van der Hucht, G. Koenigsberger, \& P.R.J. Eenens (eds.), 
        IAU Symp. 193, (San Francisco: ASP), in press (astro-ph/9812357)

\bibitem[]{}
Schaerer D., Contini T., Kunth D., 1999a, \aap, 341, 399

\bibitem[\protect\astroncite{Schaerer \etal}{1997}]{SCHAERERetal97}
Schaerer D., Contini T., Kunth D., Meynet G., 1997, \apj, 481, L75
(SCKM97)

\bibitem[]{}
Schaerer D., Contini T., Pindao, M., 1999b, \aas, in press
(astro-ph/9812347)

\bibitem[Schaerer & Vacca 1998]{1998ApJ...497..618S} Schaerer, D., Vacca, 
W. D. 1998, \apj, 497, 618 
 
\bibitem[\protect\astroncite{Stasi\'nska \& Leitherer}{1996}]{SL96}
Stasi\'nska G., Leitherer C. 1996, ApJS 107, 66 (SL96)

\bibitem[]{}
Stasi\'nska G., Schaerer, D. 1999, \aap, submitted

\bibitem[\protect\astroncite{Vacca \& Conti}{1992}]{VC92}
Vacca W.D., Conti P.S., 1992, \apj,401, 543 (VC92)

\bibitem[]{}
Vanzi, L., Rieke, G.H., Martin, C.L., Shields, J.C., 1996, \apj, 466, 150

\bibitem[\protect\astroncite{Walsh \& Roy}{1987}]{W-R87}
Walsh J.R., Roy J.-R., 1987, \apj, 319, L57 (WR87)

\bibitem[\protect\astroncite{Walsh \& Roy}{1989}]{W-R89}
Walsh J.R., Roy J.-R., 1989, \mnras, 239, 297 (WR89)

\bibitem[]{}
Walsh J.R., Roy J.-R., 1993, \mnras, 262, 27 (WR93)

\end{thebibliography}
\end{document}